\RequirePackage{fix-cm}
\documentclass[10pt,conference]{IEEEtran} 
\usepackage{tcolorbox}
\usepackage{fontawesome5}
\usepackage{enumitem}
\usepackage{mdframed}
\usepackage{listings}
\usepackage{balance}
\usepackage{hyperref}
\usepackage{cleveref}
\usepackage{booktabs}
\usepackage{multirow}
\usepackage[table]{xcolor} 
\usepackage{xfp}
\usepackage{todonotes}
\newcommand{\etal}{\textsl{et al.}\xspace}

\newcommand{\RQ}[2]{
	\vspace{5pt}
	\noindent\colorbox{yellow!08}{
        \parbox{0.95\columnwidth}{
		  \textbf{\faSearch \ RQ$_{#1}$.} \emph{#2}
		}
	}
	\vspace{5pt}
}

\newcommand{\prompt}[1]{
	\vspace{5pt}
	\noindent\colorbox{blue!05}{
        \parbox{0.95\columnwidth}{
		  \textsc{}{#1}
		}
	}
	\vspace{5pt}
}

\newcommand{\geribox}[1]{   
    \begin{mdframed}[hidealllines=true, backgroundcolor=green!10]
        {\textit{#1}}
    \end{mdframed}
}

\lstdefinelanguage{Solidity}{
  ndkeywords={msg, sender, owner},
  ndkeywordstyle=\color{black},
  comment=[l]{//},
  commentstyle=\color{gray}\itshape,
  stringstyle=\color{red},
  basicstyle=\ttfamily,
  sensitive=true,
  morestring=[b]"
}

\usepackage{graphicx}
\usepackage{stfloats}
\usepackage{float}
\usepackage{float}
\usepackage{xspace}
\begin{document}
\title{When LLM Architectures Meet Smart Contract Vulnerability Detection: a Replication Study}

\author{
\IEEEauthorblockN{Gerardo Iuliano}
\IEEEauthorblockA{University of Salerno, Italy\\
geiuliano@unisa.it}
\and
\IEEEauthorblockN{Gabriele De Vito}
\IEEEauthorblockA{University of Salerno, Italy\\
gadevito@unisa.it}
\and
\IEEEauthorblockN{Dario Di Nucci}
\IEEEauthorblockA{University of Salerno, Italy\\
ddinucci@unisa.it}
}

\maketitle


\begin{abstract}
Smart contracts are immutable programs deployed on blockchain platforms, where vulnerabilities can translate directly into financial loss. Among the techniques proposed for their detection, recent work has relied increasingly on Large Language Models (LLMs) guided by carefully crafted prompts. However, the evidence accumulated so far rests almost entirely on the dense Transformer architecture behind the GPT line and on model snapshots already scheduled for deprecation, leaving open the question of whether the reported behaviors reflect a general property of LLMs or an artifact of that specific design.

This paper presents an extended replication of the empirical study by Chen et al. on LLM-based smart contract vulnerability detection. We replicate the original protocol on GPT-4 as a baseline, followed by a differentiated replication, executing the protocol on 
\textcolor{black}{representative models based on Mixture-of-Experts and RNN-like designs, together with an attempted preliminary evaluation of State Space Models.} The exact replication confirms the findings of Chen et al., including the high-recall, low-precision detection pattern they report on GPT-4. 
\textcolor{black}{Nevertheless, the differentiated replication shows that this pattern does not transfer uniformly across the evaluated models. Those from the same architectural family may also exhibit different behaviors, indicating that architectural family alone is insufficient to predict the resulting detection profile.} We observe the same qualitative dense-Transformer profile in both GPT-4 and Gemma-3, despite a gap of roughly two orders of magnitude in parameter counts, whereas two Mixture-of-Experts models from different lineages exhibit qualitatively divergent behaviors, indicating that architectural family alone is insufficient to predict the detection profile.
\end{abstract}


\section{Introduction}
\label{sec:introduction}
A smart contract is a self-executing program that runs on a blockchain platform, such as Ethereum~\cite{buterin_ethereum}. It automatically enforces and executes the terms of an agreement when predefined conditions are met, without the need for intermediaries. Unlike traditional software, once deployed, smart contracts cannot be modified or patched, so even minor flaws can lead to severe repercussions~\cite{attacks}.

Ensuring the security of smart contracts is critical. Well-known vulnerabilities, including Reentrancy, Timestamp Dependence, Transaction Order Dependence, Authorization via \texttt{tx.origin}, and Unchecked External Calls, have already led to major financial losses, most notably in the DAO hack~\cite{dao_attack}.
To address these issues, static and dynamic analysis techniques have long been used by researchers, as extensively surveyed by Rameder et al.~\cite{ramederSLR} and Iuliano et al.~\cite{iulianoSLR}. Recently, the focus has shifted to methods based on Large Language Models (LLMs), including VulnHunt~\cite{vulnhunt}, GPTScan~\cite{gptscan}, AdvSCanner~\cite{advscanner}, LLMSmartSec~\cite{llmsmartsec}, GPTLens~\cite{gptlens}, FELLMVP~\cite{fellmvp}, and SolGPT~\cite{solgpt}, demonstrating the ability of LLMs to discover vulnerabilities by leveraging carefully designed prompts.

Within this line of work, the empirical study by Chen et al.~\cite{chen2025chatgpt} stands out as a reference point. They systematically evaluated three GPT model snapshots on the SmartBugs curated benchmark~\cite{ferreira2020smartbugs}, reporting high recall coupled with low and unstable precision. However, two aspects of their setup leave open questions worth investigating. First, their investigation focuses on a single architectural family, i.e., the dense Transformer used in the GPT line, leaving it unknown whether the observed behavior is a property of LLMs in general or an artifact of that design. Second, the study relies on model snapshots scheduled for deprecation by OpenAI~\cite{openai2023deprecation}, which threatens the long-term reproducibility and practical relevance of their conclusions. Meanwhile, the architectural landscape of LLMs has diversified considerably, with Mixture-of-Experts, RNN-like, and State Space Model designs now competing with the dense Transformer at comparable scales.

This paper presents an extended replication of Chen et al.~\cite{chen2025chatgpt} to verify whether their findings hold under a change in the underlying architecture. We organize the work around two research questions. The first asks whether the original results can be reproduced under controlled conditions on GPT-4 itself (exact replication). The second question asks whether the same protocol, prompt, dataset, and evaluation procedure yield the same behavior when the model is drawn from a different architectural family (differentiated replication). Therefore, we re-execute the experiment on representative models from three further families: Mixture-of-Experts, RNN-like, and, in a preliminary capacity, State Space Models. 
\textcolor{black}{By keeping the prompt, dataset, evaluation procedure, and temperature fixed, we examine whether the behavior observed for GPT-4 transfers to a set of models with heterogeneous architectural designs. However, because the selected models also differ in scale, training data, tokenization, tuning, and cut-off, our design does not isolate architecture as the only causal factor.}

The results confirm Chen et al.'s findings~\cite{chen2025chatgpt} in the exact replication, including the high-recall, low-precision pattern observed for GPT-4. However, the differentiated replication shows that this pattern does not generalize uniformly across architectures. In particular, the same qualitative dense-Transformer profile is observed in both GPT-4 and Gemma-3 despite a substantial scale gap, whereas DeepSeek-V2-Lite and RWKV-7 exhibit much more conservative behavior, achieving higher specificity at the cost of substantially lower recall. At the same time, the contrast between Qwen-3.5 and DeepSeek-V2-Lite shows that shared membership in the same architectural family is not, by itself, sufficient to predict detection behavior.

\textbf{Paper Structure.} \Cref{sec:background} reviews related work and key concepts to understand the paper, whereas \Cref{sec:ref_paper} describes the reference work by Chen et al.~\cite{chen2025chatgpt}.
\Cref{sec:method} details the methods used to conduct the study, with its execution plan and procedures to ensure transparency and reproducibility, and whose results are reported in Sections \ref{sec:exact} and \ref{sec:differentiated}, and discussed in \Cref{sec:discussion}.
\Cref{sec:ttv} lists the threats to the validity of the study, addressing potential limitations and biases in the study and the measures taken to mitigate them. Finally, \Cref{sec:conclusion} concludes the paper.

\section{Background and Related Works}
\label{sec:background}
This section provides background on smart contracts, their vulnerabilities, an introduction to large language models and their architectures, and related work.

\subsection{Smart Contracts and their Vulnerabilities}
Nick Szabo introduced smart contracts in 1994~\cite{szabo1994smartcontract}, but they have found new life on the Ethereum blockchain. They are self-executing programs that automatically enforce agreements between untrusted parties~\cite{antonopoulos2018masteringSc, 9667515}. Initially intended to automate legal contracts, they have evolved with the emergence of blockchain technology, becoming scripts that execute consistently across all nodes in a distributed ledger~\cite{8847638}. Developers write smart contracts in Solidity~\cite{dannen2017introSolidity} or Vyper~\cite{vyper}, two languages explicitly designed for this purpose.

In Ethereum, smart contracts execute within the Ethereum Virtual Machine (EVM), a Turing-complete, stack-based virtual machine that ensures the isolation of contract code. Each smart contract has a unique address, its own private storage, an Ether balance, and executable code. When a transaction is sent to a contract address, it activates the contract's functions, using the provided input data and consuming gas to cover computation costs~\cite{9667515}. Once deployed on the blockchain, these contracts become immutable and tamper-proof~\cite{kaushal2021immutable, politou2019blockMutability}, automatically executing predefined conditions to ensure trust and reliability.
Nevertheless, immutability, while ensuring transparency and trust, makes code defects permanent~\cite{politou2019blockMutability}, making smart contracts susceptible to a range of security vulnerabilities stemming from coding errors and software development bugs. Among the most critical are the following:

\begin{itemize}

    \item \textit{Reentrancy (SWC-107)}~\cite{Rameder2022, vidal2024openscv, li2024static, iuliano2026smart} is one of the most dangerous vulnerabilities, famously exploited in the DAO hack~\cite{kushwaha, reguard}. It occurs when a contract invokes an external contract before updating its own state, thereby enabling recursive calls that can drain funds or disrupt the intended execution flow. 
    
    \item \textit{Timestamp Dependency (SWC-116)}~\cite{Rameder2022, vidal2024openscv, li2024static, iuliano2026smart} occurs when a contract relies on the \textit{block.timestamp} value to make critical decisions, such as triggering payouts, closing auctions, or generating randomness. Since miners have limited control over the block timestamp within a reasonable range, it can be exploited to manipulate contract behavior~\cite{mense, contractfuzzer}, especially in scenarios with small time windows or financial incentives. 
    
    \item \textit{Integer Overflow and Underflow (SWC-101)}~\cite{Rameder2022, vidal2024openscv, li2024static, iuliano2026smart} occur when arithmetic operations on integer variables exceed their maximum or minimum representable limits. These bugs can lead to unexpected behavior, including bypassing balance checks or incorrectly locking or unlocking tokens. Starting with Solidity 0.8.0, the compiler automatically reverts on overflow or underflow, helping prevent such vulnerabilities by default. 
    
    \item \textit{Transaction-Ordering Dependence (TOD)} creates unpredictability in transaction execution order, which attackers can exploit through front-running to alter the outcome of a pending transaction~\cite{8726833, sayeed, etherfuzz}. 
    
    \item \textit{Denial of Service (DoS)} vulnerabilities can arise from excessive gas usage in external calls or from loops that consume excessive computation, rendering a contract temporarily or permanently inoperable~\cite{smartscan}.

\end{itemize}

\subsection{Large Language Models and their Architecture}
Large Language Models (LLMs) are advanced AI systems that can understand and generate human-like text~\cite{4} and generalize across a wide range of tasks~\cite{5,6,de2025llm}. 
They are trained on vast amounts of text data, and use deep learning with billions, sometimes trillions, of parameters to learn patterns in language. Three properties usually describe an LLM: the number of parameters, which reflects how much the model has learned to store; the context window, which bounds how much text it can consider at once; and the architecture, which determines how each token is actually processed inside the network.
Below, we present a non-exhaustive taxonomy of the main architectures of modern LLMs that we employ in the study.
Two aspects govern what happens when a token enters a model: (i) which parameters are activated, and (ii) which other tokens it can attend to.

\begin{itemize}

    \item In dense Transformers, all parameters participate in processing every token, and self-attention allows each token to interact with any other token in the input. This design underlies models such as GPT-4~\cite{achiam2023gpt}, Gemini~\cite{2025gemini}, LLaMA~\cite{touvron2023llama}, and Claude~\cite{anthropic2023claude}, explaining both their expressive power and their computational cost, which scales quadratically with input length.
    
    \item Mixture of Experts (MoE)~\cite{jiang2024mixtral} models divide their parameters into specialized subnetworks known as experts, in contrast to dense Transformers, and only activate a small subset of these subnetworks for each input token. While keeping a high total parameter count, this selective activation lowers effective compute per token. As a result, two MoE models of similar scale may behave quite differently, as performance depends more on routing policies and expert specialization than on raw size.
    
    \item RNN-like architectures--such as RWKV~\cite{peng2023rwkv}--do not include an attention mechanism. The input history is compressed into a compact recurrent state, which is updated at each token-processing stage. For long sequences, this yields linear-time processing, but at the expense of less direct access to earlier tokens, which are available only as long as they remain in the state.
    
    \item State Space Models (SSMs) occupy a middle ground between RNNs and Transformers, and are an active research direction~\cite{dao2024transformers,ghattas2025pruning,zhanrouting}. They do not compute attention over all previous tokens but maintain a fixed-sized hidden state that is updated recurrently as each token is processed, serving as a compressed memory of the sequence seen so far. Unlike attention-based models, the model has no direct access to earlier tokens, and therefore it can handle long sequences more efficiently~\cite{patro2025mamba}.

\end{itemize}

Beyond these architectures, the LLM landscape includes further variants, encoder-only Transformers~\cite{7}, diffusion-based language models~\cite{nie2026large}, sub-quadratic attention schemes~\cite{tay2022efficient}, and reasoning models~\cite{jaech2024openai,guo2025deepseek} that incorporate multi-step deliberation directly into the generation process.

\subsection{Related Work}
Boi et al.~\cite{10.1145/3687251.3687253} propose an approach for smart contract vulnerability detection by fine-tuning relatively lightweight LLMs, specifically Llama-2-7b-chat-hf and GPT-2-XL, on a curated dataset of known vulnerabilities. The fine-tuning process adapts these models to identify specific vulnerable code segments and provide targeted remediation strategies, such as recommending the SafeMath library for arithmetic issues. The fine-tuned Llama-2-7b-chat-hf achieved 82\% accuracy in identifying problematic contracts, significantly outperforming GPT-2-XL's 62\% on the same test set.

Ding et al.~\cite{ding2025smartguard} propose SmartGuard, an LLM-enhanced framework for smart contract vulnerability detection that combines semantic code retrieval, Chain-of-Thought (CoT) prompting, and in-context learning (ICL). SmartGuard retrieves the top-k code snippets most semantically similar to a given test function using CodeBERT embeddings, generates corresponding reasoning chains through a self-check architecture that leverages LLMs' self-correcting capabilities, and assembles them into a structured prompt for vulnerability inference. The self-check mechanism iteratively prompts the model to revise incorrect predictions and, when the rethinking limit is exceeded, guides it to construct a reasoning chain from a known correct label, thereby maximizing the yield of high-quality CoTs from limited labeled data. Evaluated on the SolidiFI benchmark using GPT-3.5 Turbo, ChatGLM3-6B, and Qwen-7B, SmartGuard outperforms classical static analysis tools such as Slither, Securify, and Mythril.

Wei et al.~\cite{wei2024llm} propose LLM-SmartAudit, a multi-agent conversational framework designed to automate smart contract security auditing using LLMs. Rather than relying on a single model query, the system orchestrates a team of specialized agents, a Project Manager, a Smart Contract Counselor, an Auditor, and a Solidity Programming Expert, which alternate between assistant and user roles, collaborating through structured conversations to analyze contracts, identify vulnerabilities, and produce comprehensive audit reports. The framework operates in two complementary modes: a Broad Analysis (BA) mode, based on a Thought-Reasoning prompt, and a Targeted Analysis (TA) mode, based on a Buffer-Reasoning prompt. Evaluated against ten traditional tools on a labeled dataset of 110 annotated contracts spanning ten vulnerability categories, LLM-SmartAudit in BA mode using GPT-3.5-turbo outperforms all baselines. On a real-world dataset of 6,454 contracts drawn from Code4rena audit contests, TA mode further achieves a recall of 47.55\% on complex logic vulnerabilities, a category where traditional tools score zero. 

\subsection{Research Gap}
While the aforementioned works advance smart contract vulnerability detection through fine-tuning~\cite{10.1145/3687251.3687253}, retrieval-augmented prompting~\cite{ding2025smartguard}, and multi-agent frameworks~\cite{wei2024llm}, they share a common limitation: none systematically investigates whether behavioral findings observed for a specific LLM remain consistent across changes in the underlying architecture. Boi et al.~\cite{10.1145/3687251.3687253} focus on comparing individual model performance within a fine-tuning paradigm without examining cross-architectural generalization. Despite evaluating across multiple model families, SmartGuard~\cite{ding2025smartguard} attributes its performance gains to the design of its retrieval-augmented pipeline rather than to the intrinsic transferability of vulnerability-reasoning capabilities across architectures. Similarly, LLM-SmartAudit~\cite{wei2024llm}, while providing evidence that performance varies across prompting strategies and GPT model generations, remains confined to the GPT family of architectures. Our work addresses this gap by systematically replicating and extending the findings of Chen et al.~\cite{chen2025chatgpt} across heterogeneous LLM architectures, without altering the detection methodology, to determine whether the reported behavioral patterns are model-specific or architecture-agnostic.
\section{Reference Work}
\label{sec:ref_paper}

Chen et al.~\cite{chen2025chatgpt} conducted an empirical study to investigate the effectiveness of GPT models in detecting vulnerabilities in Solidity smart contracts, evaluating three GPT model variants on the widely used smartbugs-curated dataset~\cite{ferreira2020smartbugs}.

The authors interacted with the GPT models via the official OpenAI API~\cite{openai2023api}, targeting three specific model snapshots: gpt-3.5-turbo-0613~\cite{openai2023gpt35}, gpt-4o-2024-05-13~\cite{openai2024gpt4o}, and gpt-4-0613~\cite{openai2023gpt4}. The temperature parameter was set to 0.7 across all experiments to balance creativity and stability in text generation, while all remaining parameters were kept at their default values (e.g., TopP = 1, FrequencyPenalty = 0).
A central methodological contribution of the study is its prompt design process. Models based on GPT architecture rely on a prompt-based learning approach~\cite{liu2023pre}. Rather than manually crafting a prompt, the authors employed GPT-4 itself as a prompt optimizer~\cite{lo2024large}. Starting from a naive task description, they fed the initial prompt into the model and iteratively refined it. Multiple draft prompts were generated and manually evaluated on a sample of 20 contracts drawn from the smartbugs-curated dataset. This process converged on the following general-purpose template:

\prompt{\textbf{Prompt Template:} \textit{You are [ROLE]. [TASK DESCRIPTION]. Think step by step, carefully. The input is [INPUT].}}

Two distinct roles were instantiated within this template. The first, designated as a smart contract vulnerability detector, was used to perform the actual detection task. The prompt was filled as follows:

\prompt{\textbf{Detection Prompt:} \textit{You are a vulnerability detector for a smart contract. Here are nine
common vulnerabilities. [VULS]. Check the following smart contract
for the above vulnerabilities. Think step by step, carefully.
The input is [INPUT].}}

Placeholder [VULS] was replaced with the names and aliases of the nine DASP Top 10 vulnerability categories, and [INPUT] with the source code of the contract under analysis.

The second role, designated as text semantic analyzer, was invoked in a separate conversation session to normalize the free-form output of the first step into a structured binary format ([VUL]: 0/1). This two-step pipeline was motivated by an empirical finding reported by the authors: directly constraining the output format within the detection prompt degraded performance across most vulnerability types (Finding 1). The semantic analysis prompt took the following form:

\prompt{\textbf{Formatting Prompt:} \textit{You are a semantic analyzer of text. Here are nine common
vulnerabilities. [VULS]. The following text is a vulnerability
detection result for a smart contract. Use 0 or 1 to indicate
whether there are specific types of vulnerabilities. For example:
"Reentrancy: 1". Think step by step, carefully. The input is [INPUT].}}

To validate the reliability of the semantic analysis step, the authors manually inspected the normalized output for 50 sampled contracts. Of these, 47 were entirely consistent with the original detection output, 2 exhibited omissions, and 1 showed hallucinations, confirming the overall soundness of the approach.
Finally, to account for the known non-determinism of GPT models, each experiment was repeated five times on the full dataset, and the standard deviation of the F1 score across runs was reported alongside precision, recall, specificity, and F1.
\section{Research Method}
\label{sec:method}

\begin{figure*}[ht]
    \centering
    \includegraphics[width=1\linewidth]{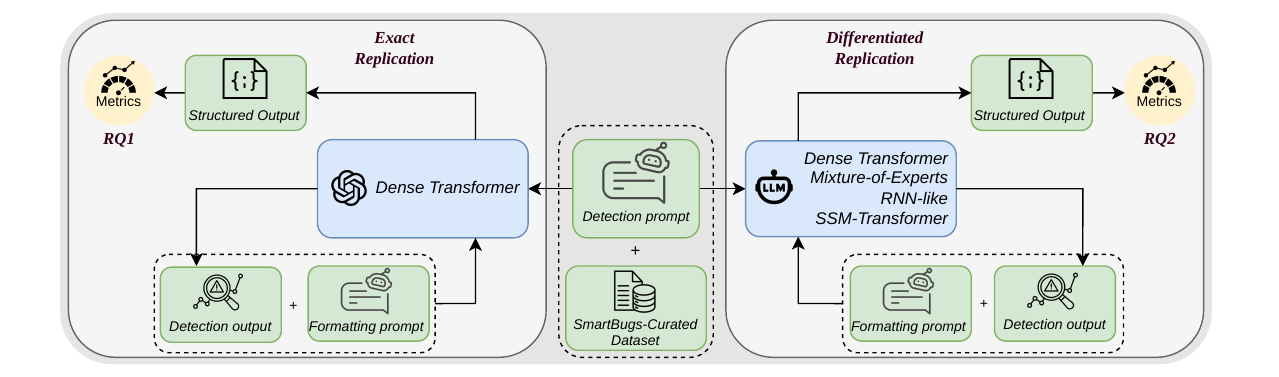}
    \caption{Research method we followed to conduct the exact replication and the differentiated replication.}
    \label{fig:method}
\end{figure*}

The empirical study by Chen et al.~\cite{chen2025chatgpt} relies on three specific GPT model snapshots scheduled for deprecation by OpenAI and scheduled for shutdown in 2026~\cite{openai2023deprecation}, raising an important concern about the longevity and generalizability of their findings. If the results are tightly coupled to a specific architectural family that is becoming unavailable, their practical applicability is inherently limited. We therefore investigate whether the reported findings hold across different LLM architectural families or are an artifact of the GPT-4 architecture. \Cref{fig:method} shows the research method we followed. Specifically, the exact replication is represented on the left, while the differentiated replication is depicted on the right.

To achieve our goal, we formulated the following research questions.

\RQ{1}{Can the results reported in the original study be replicated using the same prompt, dataset, and method?}

$RQ_1$ establishes the baseline of the replication. It verifies that the original findings are stable and reproducible under controlled conditions with the same prompt, dataset, and metrics, before applying changes, thereby revealing discrepancies arising from model versioning, non-determinism, or methodological ambiguity in the original study.

\RQ{2}{Do the findings of Chen et al. hold across different LLM architectural families?}

\textcolor{black}{$RQ_2$ assesses transfer across selected models representing different architectural families without estimating the isolated causal effect of architecture.}
The evaluation conducted by Chen et al.~\cite{chen2025chatgpt} relied on GPT-4, a dense Transformer model. $RQ_2$  investigates whether their main findings, namely, high recall at the cost of low precision and substantial performance variation across vulnerability types, hold across different LLM architectural families.





\subsection{Dataset}

As in the original study~\cite{chen2025chatgpt}, we used the SmartBugs curated dataset~\cite{ferreira2020smartbugs}, which comprises 142 vulnerable smart contracts across nine vulnerability types described in the DASP Top 10~\cite{wong2018dasp}.
Smart contracts are annotated with a comment that lists the line numbers at which vulnerabilities are reported, along with their types. As in the original paper, we removed all comments.

\subsection{Execution and Analysis}
All experiments were conducted using the same prompts as in the reference paper and were repeated five times to mitigate model instability. To evaluate the detection results, as in the previous paper~\cite{chen2025chatgpt}, we used precision, recall, F1 score, and specificity. We also calculated the standard deviation of the F1-score to quantify performance variability and evaluate the consistency of the results.
To make the recall--false-positive trade-off easier to read, we used Youden's J statistic~\cite{ruopp2008youden}, defined as $J = \mathrm{TPR} - \mathrm{FPR}$, or equivalently $J = \mathrm{Sensitivity} + \mathrm{Specificity} - 1$. Since Recall, TPR, and Sensitivity coincide in this setting, $J$ summarizes in a single number how much a model improves over random guessing while balancing its ability to recover true positives against its tendency to raise false alarms. Higher values indicate a better trade-off, values near zero indicate weak discrimination, and negative values mean that false-positive behavior outweighs true-positive recovery.

\subsection{Data Availability}
The datasets and results of this paper are publicly available at \url{https://figshare.com/s/bbb79bd8002deb1d40ac}.

\section{RQ$_1$ - Exact Replication}
\label{sec:exact}

\subsection{RQ$_1$ Design}
The reference study~\cite{chen2025chatgpt} showed that the GPT-4 model~\cite{openai2023gpt4} has the best performance compared to GPT-3.5-turbo and GPT-4o.
To answer $RQ_1$, we repeated the experiment using the same model, GPT-4-0613, with the same settings: a temperature of 0.70 and all other parameters at their default values. 
Similarly, we processed the output with a second prompt rather than enforcing strict formatting in the initial response, given the limitations of GPT in consistently adhering to such formats~\cite{chen2025chatgpt}.
This two-step approach leverages the semantic understanding of the model, improves its reliability, ensures standardized output (e.g., ``[VUL]: 0/1''), and reduces the risk of formatting errors during detection~\cite{chen2025chatgpt}.
We relied on the information provided in the paper and its replication package.

\subsection{RQ$_1$ Results}

\begin{table*}[ht]
\centering
\scriptsize
\caption{Performance metrics obtained by replicating the original study using the same prompt, dataset, and method.}
\label{tab:comparison}
\rowcolors{2}{gray!10}{white}
\begin{tabular}{l|ccccc|cccccc}
\toprule
\multirow{2}{*}{\textbf{Vulnerability}} 
& \multicolumn{5}{c|}{\textbf{Chen et al. Results}} 
& \multicolumn{6}{c}{\textbf{Our Results}} \\
\cmidrule(lr){2-6} \cmidrule(lr){6-12}
& \textbf{Precision} & \textbf{Recall} & \textbf{Specificity} & \textbf{F1-score} & \textbf{F1 Std.} 
& \textbf{Precision} & \textbf{Recall} & \textbf{Specificity} & \textbf{F1-score} & \textbf{F1 Std.} & \textbf{Youden’s J} \\
\midrule
Access Control             & 0.17 & 0.96 & 0.30 & 0.28 & 0.04 & 0.17 & 0.98 & 0.33 & 0.30 & 0.01 & 0.31 \\
Arithmetic                 & 0.22 & 0.96 & 0.63 & 0.36 & 0.03 & 0.24 & 0.92 & 0.66 & 0.38 & 0.03 & 0.58 \\
Bad Randomness             & 0.40 & 0.98 & 0.91 & 0.57 & 0.04 & 0.46 & 0.95 & 0.94 & 0.63 & 0.03 & 0.88 \\
Denial of Service          & 0.07 & 0.97 & 0.42 & 0.13 & 0.02 & 0.07 & 1.00 & 0.38 & 0.12 & 0.01 & 0.38 \\
Front Running              & 0.09 & 0.70 & 0.81 & 0.16 & 0.01 & 0.12 & 0.50 & 0.88 & 0.19 & 0.09 & 0.38 \\
Reentrancy                 & 0.34 & 0.97 & 0.41 & 0.50 & 0.03 & 0.40 & 0.97 & 0.57 & 0.56 & 0.01 & 0.54 \\
Short Addresses            & 0.05 & 0.60 & 0.91 & 0.08 & 0.00 & 0.06 & 0.80 & 0.91 & 0.11 & 0.06 & 0.71 \\
Time Manipulation          & 0.17 & 0.88 & 0.83 & 0.28 & 0.04 & 0.12 & 1.00 & 0.72 & 0.21 & 0.01 & 0.72 \\
Unchecked Calls  & 0.54 & 0.93 & 0.68 & 0.68 & 0.05 & 0.49 & 0.95 & 0.42 & 0.65 & 0.01 & 0.37 \\
\midrule
\rowcolor{white}
\textbf{Macro Average}     & 0.23 & 0.88 & 0.66 & 0.34 & 0.03 & 0.24 & 0.90 & 0.65 & 0.35 & 0.03 & 0.54 \\
\textbf{Micro Average}     & --   & --   & --   & --   & --   & 0.26 & 0.93 & 0.68 & 0.41 & --   & 0.61 \\
\bottomrule
\end{tabular}
\end{table*}


The side-by-side comparison of our replication and the results reported by Chen \etal~\cite{chen2025chatgpt} provides insights into both the stability and the variability of GPT-4's performance on the smartbugs-curated dataset. \Cref{tab:comparison} shows both results. At a high level, the two experiments confirm a similar trend: GPT-4 achieves consistently high recall across most vulnerability classes but exhibits relatively low precision, resulting in modest F1 Scores. 
\textcolor{black}{Although the dataset, prompt, model snapshot, and reported decoding configuration were kept consistent with the reference study, our replication highlights several important differences, suggesting that residual differences may arise from inference non-determinism, undocumented provider-side implementation details, API behavior, or ambiguities in the original execution and post-processing procedures.}

First, recall is robust across both studies, often close to or at $1.00$ for classes such as denial-of-service and time manipulation, indicating that GPT-4 systematically favors sensitivity over selectivity, detecting almost all true vulnerabilities at the cost of many false positives. Interestingly, while the original study reported slightly lower recall in denial-of-service (0.97 vs. 1.00 in our results), our replication shows higher recall for time manipulation (1.00 vs. 0.88). 
\textcolor{black}{Such discrepancies indicate that class-level estimates remain sensitive to stochastic generation and potentially undocumented execution details, even when the published protocol is closely reproduced.}

Second, precision exhibits a more pronounced divergence between the two studies. For example, in reentrancy, our results yield higher precision (0.40 vs. 0.34), while in unchecked low-level calls, the original study reports stronger precision (0.54 vs. 0.49). Similarly, in bad randomness, our replication produces higher F1-scores (0.63 vs. 0.57). These shifts illustrate that, although the model's overall tendency is consistent, its reliability in identifying specific vulnerability classes is variable, even after five rounds of execution.

Third, specificity values underscore an important trade-off. In some categories (e.g., short addresses), both studies agree on relatively high specificity. However, for others, the results diverge, i.e., front running shows lower specificity in the original (0.81) compared to our replication (0.88). These differences suggest that GPT-4's balance between FPs and false negatives is unstable across replications.

Fourth, the FPR values reported in our replication further confirm GPT-4's tendency to over-predict vulnerabilities. The macro- and micro-averages of 0.36 and 0.32, respectively, indicate that approximately one in three non-vulnerable contracts is incorrectly flagged. This behavior is particularly pronounced for access control (FPR = 0.67) and denial-of-service (FPR = 0.62), where the model raises a false alarm in most negative cases. Similarly, unchecked low-level calls (FPR = 0.58) and reentrancy (FPR = 0.43) exhibit higher FPRs, suggesting that GPT-4 struggles to distinguish vulnerable patterns from benign ones in these categories. In contrast, bad randomness (FPR = 0.07) and short addresses (FPR = 0.09) stand out as the classes in which the model produces the fewest false alarms, likely reflecting more pronounced syntactic or semantic patterns. These FPR disparities align with the low precision observed across most categories and reinforce the concern that GPT-4, while rarely missing a true vulnerability, does so at the expense of a high false-positive rate.

Looking at the macro averages, both sets of results are closely aligned: precision remains at 0.23--0.24, recall between 0.88--0.90, and F1-score at 0.34--0.35. This consistency indicates that, although class-level variations exist, the aggregate performance of GPT-4 is relatively reproducible.

\geribox{
\textbf{RQ$_1$ Summary}. Our replication broadly validates the conclusions of the reference study but highlights significant variability across vulnerability classes. GPT-4 achieves high recall, making it suitable when missing vulnerabilities are unacceptable. Nevertheless, its low and unstable precision, coupled with a macro-average FPR of 0.36, raises concerns about the number of FPs, thereby hindering the practical applicability of the method.
}
\begin{table}[ht]
\centering
\caption{LLM architectural families and representative models selected for RQ2.}
\label{tab:rq2_models}
\rowcolors{2}{gray!10}{white}
\scriptsize
\begin{tabular}{l|l|r|r|r}
\toprule
\textbf{Architectural Family} & \textbf{Model} & \textbf{Total} & \textbf{Active} & \textbf{Context} \\
\midrule
Dense Transformer      & Gemma 3 12B    & 12B   & 12B   & 128K \\
Mixture of Experts     & Qwen 3.5 35B a3b   & 35B   & 3B    & 262K \\
                       & DeepSeek-V2-Lite   & 16B & 2.4B & 32K  \\
RNN-like               & RWKV-7 13.3B   & 14B   & 14B   & 1M   \\
\bottomrule
\end{tabular}
\end{table}

\section{RQ$_2$ - Differentiated Replication}
\label{sec:differentiated}

\subsection{RQ$_2$ Design}
\Cref{tab:rq2_models} reports the LLM architectural families and the selected representative model.

The dominant architecture in modern LLMs is the dense Transformer decoder, which underpins GPT-4. To assess whether the findings of Chen et al. are attributable to this architectural choice or reflect a more general behavior, we selected one representative model per major architectural family, keeping the parameter count comparable across models where possible to avoid conflating architectural effects with scale effects. Reasoning models were excluded from this analysis, because the original prompt (see RQ$_1$) already incorporates a Chain-of-Thought pattern, and applying it to a reasoning model would conflate the effect of the prompt pattern with the model's internal mechanism, making it impossible to attribute observed performance differences to either factor in isolation.

Gemma 3 12B was selected as the dense Transformer reference point because it belongs to the same architectural family as GPT-4. Qwen 3.5 35B a3b represents the MoE family. We include a second MoE representative, DeepSeek-V2-Lite 16B, whose total parameter count is approximately 16 billion with roughly 2.4 billion active per token. The reason for including two MoE models is that Qwen 3.5 and DeepSeek-V2-Lite come from different development lineages and differ in their expert routing strategies, allowing us to distinguish behavior that is attributable to the MoE mechanism from behavior that is vendor-specific.

Finally, RWKV-7 14B is an RNN-like architecture and is the most structurally distant model from GPT-4 in our selection.

The rationale for this selection is architectural diversity rather than performance maximization. We are not asking which model performs best, nor whether the prompt transfers effectively. We are asking whether the empirical pattern reported by Chen et al., namely, high recall coupled with low precision, emerges consistently regardless of how the underlying model processes sequences, or whether it is contingent on the Transformer attention mechanism.

The two-step prompting process is maintained to ensure consistent output formatting, and all models are run at a temperature of 0.70, as in RQ$_1$.

We also included SSMs in our preliminary evaluation. Specifically, we evaluated Bamba 9B, a pure SSM, and Jamba 2 3B, a hybrid SSM-Transformer model that interleaves standard attention layers with state-space layers, sitting architecturally between the two families (i.e., SSMs and Transformers). However, both models exhibited a systematic failure mode on the detection task. Rather than producing the expected vulnerability assessment, generation entered repetitive loops that consumed the available context window without converging to a terminating output. Consequently, the outputs could not be parsed by the formatting prompt pipeline used in RQ$_1$ and RQ$_2$, and the metrics could not be computed, indicating not poor detection performance but rather that the models did not produce well-formed outputs for which precision, recall, or specificity could be defined. This failure mode is consistent with reported limitations of current SSM checkpoints in long-form instruction following~\cite{wangunderstanding, patro2025mamba}. Therefore, we report it as a negative observation rather than a quantitative comparison, and leave to future work a focused investigation of whether this failure is intrinsic to the architecture or an artifact of the current generation of instruction-tuned checkpoints.

\subsection{RQ$_2$ Results}
\begin{table*}[ht]
\centering
\scriptsize
\caption{Performance comparison across different LLM architectural families, including Youden's J statistic.}
\begin{tabular}{llccccc|c}
\toprule
\textbf{Architecture} & \textbf{Vulnerability} & \textbf{Precision} & \textbf{Recall} & \textbf{Specificity} & \textbf{F1-score} & \textbf{F1 Std.} & \textbf{Youden's J}\\
\midrule
 & Access Control            & 0.15 & 0.92 & 0.25 & 0.26 & 0.02 & 0.18 \\
 & Arithmetic                & 0.16 & 0.98 & 0.41 & 0.28 & 0.01 & 0.39 \\
 & Bad Randomness            & 0.42 & 0.88 & 0.93 & 0.57 & 0.01 & 0.81 \\
 & Denial of Service         & 0.05 & 1.00 & 0.09 & 0.09 & 0.00 & 0.09 \\
 Dense Transformer & Front Running             & 0.03 & 0.75 & 0.30 & 0.06 & 0.02 & 0.05 \\
 (Gemma-3 12B) & Reentrancy                & 0.28 & 1.00 & 0.25 & 0.44 & 0.00 & 0.25 \\
 & Short Addresses           & 0.00 & 0.00 & 0.80 & 0.00 & 0.00 & -0.20 \\
 & Time Manipulation         & 0.10 & 0.93 & 0.70 & 0.19 & 0.02 & 0.63 \\
 & Unchecked Calls           & 0.46 & 0.99 & 0.31 & 0.63 & 0.02 & 0.30 \\
\cmidrule{2-8}
 & Macro Average    & 0.18 & 0.83 & 0.45 & 0.28 & 0.01 & 0.28 \\
 & Micro Average    & 0.18 & 0.96 & 0.46 & 0.31 & ---  & 0.42 \\
\midrule
 & Access Control            & 0.16 & 0.62 & 0.51 & 0.25 & 0.03 & 0.13 \\
 & Arithmetic                & 0.23 & 0.92 & 0.65 & 0.38 & 0.03 & 0.57 \\
 & Bad Randomness            & 0.41 & 0.95 & 0.92 & 0.57 & 0.04 & 0.87 \\
 & Denial of Service         & 0.06 & 0.93 & 0.39 & 0.12 & 0.01 & 0.32 \\
 Mixture-of-Experts & Front Running             & 0.03 & 0.35 & 0.68 & 0.06 & 0.04 & 0.03 \\
 (Qwen-3.5 35B a3b) & Reentrancy                & 0.34 & 0.99 & 0.45 & 0.51 & 0.01 & 0.44 \\
 & Short Addresses           & 0.02 & 0.20 & 0.88 & 0.04 & 0.09 & 0.08 \\
 & Time Manipulation         & 0.15 & 0.84 & 0.82 & 0.25 & 0.03 & 0.66 \\
 & Unchecked Calls           & 0.67 & 0.99 & 0.71 & 0.80 & 0.03 & 0.70 \\
\cmidrule{2-8}
 & Macro Average    & 0.23 & 0.75 & 0.67 & 0.33 & 0.03 & 0.42 \\
 & Micro Average    & 0.26 & 0.90 & 0.67 & 0.40 & ---  & 0.57 \\
\midrule
 & Access Control             & 0.17 & 0.69 & 0.50 & 0.27 & 0.04 & 0.19 \\
 & Arithmetic                 & 0.11 & 0.13 & 0.87 & 0.12 & 0.08 & 0.01 \\
 & Bad Randomness             & 0.14 & 0.40 & 0.86 & 0.21 & 0.08 & 0.26 \\
 & Denial of Service          & 0.08 & 0.17 & 0.90 & 0.11 & 0.11 & 0.07 \\
 Mixture-of-Experts  & Front Running              & 0.02 & 0.05 & 0.92 & 0.02 & 0.05 & -0.03 \\
 (DeepSeek-V2-Lite 16B) & Reentrancy              & 0.31 & 0.51 & 0.66 & 0.38 & 0.05 & 0.18 \\
 & Short Addresses            & 0.00 & 0.00 & 0.90 & 0.00 & 0.00 & -0.10 \\
 & Time Manipulation          & 0.16 & 0.36 & 0.93 & 0.22 & 0.07 & 0.29 \\
 & Unchecked Calls            & 0.42 & 0.31 & 0.75 & 0.36 & 0.08 & 0.06 \\
\cmidrule{2-8}
 & Macro Average     & 0.15 & 0.29 & 0.81 & 0.19 & 0.06 & 0.10 \\
 & Micro Average     & 0.21 & 0.38 & 0.82 & 0.27 & ---  & 0.20 \\
\midrule
 & Access Control             & 0.14 & 0.25 & 0.78 & 0.18 & 0.10 & 0.04 \\
 & Arithmetic                 & 0.28 & 0.40 & 0.88 & 0.33 & 0.07 & 0.28 \\
 & Bad Randomness             & 0.27 & 0.33 & 0.95 & 0.29 & 0.10 & 0.28 \\
 & Denial of Service          & 0.06 & 0.27 & 0.81 & 0.10 & 0.07 & 0.08 \\
 RNN-like & Front Running              & 0.07 & 0.10 & 0.96 & 0.08 & 0.11 & 0.06 \\
 (RWKV7 13.3B) & Reentrancy               & 0.33 & 0.69 & 0.58 & 0.45 & 0.04 & 0.27 \\
 & Short Addresses            & 0.07 & 0.20 & 0.97 & 0.10 & 0.22 & 0.17 \\
 & Time Manipulation          & 0.05 & 0.12 & 0.94 & 0.07 & 0.11 & 0.06 \\
 & Unchecked Calls            & 0.48 & 0.36 & 0.77 & 0.41 & 0.09 & 0.13 \\
\cmidrule{2-8}
 & Macro Average     & 0.19 & 0.30 & 0.85 & 0.22 & 0.10 & 0.15 \\
 & Micro Average     & 0.26 & 0.40 & 0.86 & 0.32 & ---  & 0.26 \\
\bottomrule
\end{tabular}
\label{tab:architectures_performance}
\end{table*}





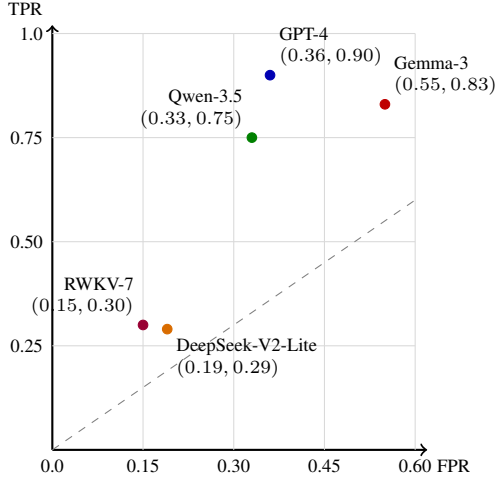
\begin{figure}[t]
\centering
\begin{tikzpicture}[x=8cm,y=5.5cm,font=\scriptsize]
    \draw[->, thick] (0,0) -- (0.62,0) node[below right] {FPR};
    \draw[->, thick] (0,0) -- (0,1.02) node[above left] {TPR};
    \draw[gray!30] (0.15,0) -- (0.15,1.0);
    \draw[gray!30] (0.30,0) -- (0.30,1.0);
    \draw[gray!30] (0.45,0) -- (0.45,1.0);
    \draw[gray!30] (0.60,0) -- (0.60,1.0);
    \draw[gray!30] (0,0.25) -- (0.60,0.25);
    \draw[gray!30] (0,0.50) -- (0.60,0.50);
    \draw[gray!30] (0,0.75) -- (0.60,0.75);
    \draw[gray!30] (0,1.00) -- (0.60,1.00);
    \draw[dashed, gray] (0,0) -- (0.60,0.60);
    \node[below] at (0,0) {0.0};
    \node[below] at (0.15,0) {0.15};
    \node[below] at (0.30,0) {0.30};
    \node[below] at (0.45,0) {0.45};
    \node[below] at (0.60,0) {0.60};
    \node[left] at (0,0.25) {0.25};
    \node[left] at (0,0.50) {0.50};
    \node[left] at (0,0.75) {0.75};
    \node[left] at (0,1.00) {1.0};
    \filldraw[blue!70!black] (0.36,0.90) circle (1.8pt);
    \node[above right, align=left] at (0.36,0.90) {GPT-4\\$(0.36,0.90)$};
    \filldraw[red!75!black] (0.55,0.83) circle (1.8pt);
    \node[above right, align=left] at (0.55,0.83) {Gemma-3\\$(0.55,0.83)$};
    \filldraw[green!50!black] (0.33,0.75) circle (1.8pt);
    \node[above left, align=right] at (0.33,0.75) {Qwen-3.5\\$(0.33,0.75)$};
    \filldraw[orange!85!black] (0.19,0.29) circle (1.8pt);
    \node[below right, align=left] at (0.19,0.29) {DeepSeek-V2-Lite\\$(0.19,0.29)$};
    \filldraw[purple!80!black] (0.15,0.30) circle (1.8pt);
    \node[above left, align=right] at (0.15,0.30) {RWKV-7\\$(0.15,0.30)$};
\end{tikzpicture}
\caption{ROC-style scatter plot using macro-average FPR (x-axis) and TPR (y-axis) for each model in RQ$_2$. Points closer to the upper-left region indicate a better balance between vulnerability recovery and false-alarm control.}
\label{fig:rq2_roc_scatter}
\end{figure}

\Cref{tab:architectures_performance} shows that findings provided by Chen et al.~\cite{chen2025chatgpt} do not uniformly generalize across different LLM architectural families. However, the degree of divergence varies substantially across architectures. 

The Qwen-3.5 35B MoE model comes closest to replicating the baseline behavior, achieving a macro F1 of 0.33 and a micro F1 of 0.40, with a macro/micro FPR of 0.33, all comparable to our replication (macro F1 0.35, micro F1 0.41, macro FPR 0.36, micro FPR 0.32). It also attains a macro Youden's J of 0.42, closer to the GPT-4 replication (0.54) than any other non-dense model, suggesting that some MoE models can approximate the dense-Transformer detection p. Still, theut the family does not do so uniformly.

In contrast, the Dense Transformer (Gemma-3 12B) maintains high recall (macro 0.83), consistent with the tendency of the baseline to flag aggressively, but at the cost of a substantially higher FPR (0.55 vs. 0.36), indicating a more indiscriminate detection strategy that inflates false positives beyond what was observed in the original study. In particular, we observe the same qualitative pattern in the two dense Transformer LLMs, which differ by a factor of two orders of magnitude in scale. Two data points are insufficient to support any claims of scale-invariance. Still, the consistency suggests that the high-recall/low-precision profile is not due solely to scale within this family of architectures. 

The most pronounced deviations occur with the DeepSeek-V2-Lite and RWKV7 models, both of which break the high-recall, low-precision pattern observed in Chen et al.'s results. These models achieve much better specificity (0.81 and 0.85, respectively) and lower FPRs (0.19 and 0.15), but at the expense of recall, which drops dramatically to 0.29 and 0.30 — less than a third of the baseline's 0.90. Their macro Youden's J values (0.10 and 0.15) are also far below those of GPT-4 and Qwen-3.5, making the trade-off evident in a single statistic. Across all models, certain vulnerability classes, such as Unchecked Low-Level Calls and Bad Randomness, remain relatively easier to detect. At the same time, Front-Running and Short Addresses consistently prove challenging regardless of architecture. Overall, the evidence suggests that the high-recall detection behavior reported by Chen et al. is not an inherent property of LLM-based vulnerability detection, but rather is sensitive to architectural choices, and practitioners should be cautious about assuming these findings transfer to smaller or non-transformer-based models.

The message is supported by the ROC-style scatter plot shown in \Cref{fig:rq2_roc_scatter}, showing that the models do not differ only in magnitude on a single metric but occupy qualitatively different operating regions. GPT-4 and Qwen occupy the upper portion of the plane, indicating that both recover a large fraction of true vulnerabilities. Still, Qwen sits slightly further to the left, indicating a somewhat better false-positive profile. Gemma-3 remains in the same high-TPR region as GPT-4 but is markedly displaced to the right, indicating its more aggressive over-flagging behavior. By contrast, DeepSeek-V2-Lite and RWKV-7 move toward the lower-left region: they reduce false alarms, but only at the cost of a substantial reduction in true-positive recovery.

The last column of \Cref{tab:architectures_performance} makes the pattern even clearer. At the macro level, GPT-4 remains the strongest reference point with a Youden's J of 0.54, followed by Qwen-3.5 at 0.42. Gemma-3 reaches only 0.28: its high recall is partially offset by a much larger false-positive burden, which is precisely why its behavior should be read as aggressive but less selective. DeepSeek-V2-Lite and RWKV-7 fall further to 0.10 and 0.15, respectively, confirming that their higher specificity comes at the cost of a sharply reduced true-positive recovery. Looking at the per-vulnerability rows, Qwen is closest to the baseline on categories such as Arithmetic, Bad Randomness, Time Manipulation, and Unchecked Calls. In contrast, DeepSeek and RWKV rarely match that balance and even become negative in isolated cases, such as Front Running or Short Addresses, where false positives outweigh true positives. In this sense, Youden's J provides a compact summary of the same message conveyed by the separate recall and specificity columns: the dense baseline and Qwen occupy a relatively effective operating region, Gemma shifts toward over-flagging, and DeepSeek-V2-Lite and RWKV-7 shift toward under-detection.


\geribox{
\textbf{RQ$_2$ Summary}. The high-recall, low-precision pattern identified by Chen et al. is not a universal characteristic of LLMs but is highly sensitive to architectural choices. While large MoE models like Qwen-3.5 closely replicate the baseline behavior, the sparse architecture of DeepSeek-V2-Lite and the RNN-like architecture of RWKV-7 13.3B exhibit a conservative shift, drastically reducing false positives but at the cost of missing most true vulnerabilities, suggesting that the original findings do not automatically transfer to these architectures.
}
\section{Discussion}
\label{sec:discussion}
The section discusses the implications of our findings for the research questions.

\subsection{Architecture, Scale, and Active Parameter Count}
A potentially important alternative explanation for our findings is scale rather than architecture. GPT-4 is vastly larger than the open models considered here, so a skeptical reader could argue that the high-recall/low-precision profile is simply what happens at sufficiently large scale, regardless of architecture. Our data do not fully rule out that possibility, but they do weaken it in one important respect. Despite being far smaller than GPT-4, Gemma-3 12B exhibits the same qualitative dense-Transformer pattern: high recall accompanied by high false-positive rates. Two data points are not enough to claim scale invariance, but they do suggest that this detection profile is not solely attributable to the dense Transformer family. \textcolor{black}{The active parameter count provides additional context for interpreting these differences. RWKV-7 activates approximately 14B parameters, a budget comparable to Gemma-3’s 12B, yet exhibits a markedly different detection profile. This comparison weakens a simple active-capacity explanation for RWKV-7, although it does not isolate architecture from differences in training data, instruction tuning, and alignment. By contrast, DeepSeek-V2-Lite activates only about 2.4B parameters per token, so limited active capacity remains a plausible contributor to its substantially lower recall.}

\subsection{MoE Architecture: Same Family, Two Philosophies}
A notable takeaway from comparing Qwen 3.5 35B a3b and DeepSeek-V2-Lite 16B is not merely that they perform differently, but that they occupy clearly different regions of the recall--false-positive trade-off despite sharing the same underlying architectural principle of sparse expert activation.

Qwen 3.5 35B a3b exhibits a consistently aggressive detection posture. Across almost every vulnerability category, its recall is notably high, indicating that it tends to flag a contract as vulnerable rather than clean. This behavior is observed across structurally distinct vulnerability types, from Reentrancy to Bad Randomness, placing Qwen relatively close to GPT-4 in ROC-style terms: high TPR paired with a comparatively elevated FPR, and a macro Youden's J of 0.42.

DeepSeek-V2-Lite 16B exhibits the opposite tendency: it is considerably more conservative, and in some categories it is almost entirely silent. Empirically, the model combines low recall with low false-positive rates, which places it in a markedly different operating region from both Qwen and GPT-4. Its macro Youden's J of 0.10 confirms that this is not simply a cleaner version of the same behavior, but a weaker discriminative profile overall.

As shown also in \Cref{fig:rq2_roc_scatter}, both models share the MoE mechanism but differ markedly in detection behavior. Qwen lies near GPT-4 in the upper-middle region of the plane, whereas DeepSeek-V2-Lite moves toward the lower-left region, making the intra-family divergence immediately apparent. Architectural family membership alone is therefore not sufficient to predict the detection profile. Differences in lineage, routing design, active parameter budget, and training signal are all plausible contributors, but our study does not isolate them individually. We can support the narrower empirical claim that two MoE models can behave qualitatively differently under the same prompting and evaluation protocol. In a security-auditing context, neither posture is inherently correct because the preferred balance depends on the cost asymmetry between false alarms and missed vulnerabilities.

\subsection{Detection Scope Limitations of RNN-like Architectures.}
RWKV-7 occupies a structurally distinct position in this comparison, as it is the only model that processes tokens sequentially via gated recurrent updates without any attention. This architectural distinction has observable consequences. Unlike every other model in the study, whether dense or sparse, RWKV-7 cannot directly relate distant tokens to one another. Instead, it compresses the preceding context into a fixed-size recurrent state at each step, meaning that information from earlier in a contract is subject to progressive decay and interference. For smart contract vulnerability detection, where meaningful signals often emerge from non-local interactions between a state variable and a function body defined hundreds of tokens apart, this is a structurally significant limitation.

The resulting detection pattern is consistently conservative. Across nearly all vulnerability categories, RWKV-7 exhibits high specificity paired with low recall, suggesting that the model rarely accumulates sufficient contextual evidence to commit to a positive detection. This behavior contrasts with Gemma-3, which aggressively flags vulnerability-suggestive patterns using global attention, sometimes indiscriminately. RWKV-7 does not overclaim; it largely remains silent but this conservatism is relaxed for categories such as Reentrancy and Unchecked Low Level Calls, where the problematic pattern tends to be locally concentrated within a single function body, consistent with what one would expect from a recurrent architecture: short-range dependencies are preserved reasonably well, while cross-function reasoning suffers.

Therefore, the inclusion of RWKV-7 in this study is analytically valuable beyond its raw performance. It serves as a structural control, isolating the attention mechanism's contribution from confounding factors such as model scale or training data. The performance gap between RWKV-7 and the attention-based models is most plausibly attributable to inductive bias rather than capacity. For tasks that demand long-range relational reasoning over structured code, attention is not merely beneficial but potentially necessary. That said, the observed scope limitations do not preclude RWKV-7 from being effective in narrower settings; function-level vulnerability detection, or any formulation in which the relevant context is compact enough to be faithfully captured within the recurrent state, may be a more natural fit for this architectural family.

\section{Threats to Validity}
\label{sec:ttv}
We discuss threats to validity following Wohlin et al.~\cite{wohlin2012experimentation}.

\paragraph*{Internal Validity}
A primary threat to internal validity concerns model non-determinism. Despite repeating each experiment five times and reporting the standard deviation of the F1-score, the stochastic nature of LLM inference at a temperature of 0.70 means that observed performance differences between architectures may partly reflect run-to-run variability rather than stable behavioral properties of the underlying model. Although our replication design mirrors the protocol of Chen et al.~\cite{chen2025chatgpt}, we cannot fully exclude that some class-level divergences are artifacts of sampling noise rather than architectural signals.

A second threat concerns the two-step prompting pipeline. While we adopted the same formatting prompt used in the reference study to normalize model outputs, different architectures may exhibit heterogeneous tendencies to deviate from the expected output schema, leading to higher rates of parsing failures or silent omissions for some models than for others. Although this design choice was motivated by the need for methodological consistency with Chen et al.~\cite{chen2025chatgpt}, it introduces a confound: differences in adherence to structured outputs across architectures may distort detection metrics independently of vulnerability reasoning capability.

A further threat relates to parameter comparability across models. Despite our effort to select models of comparable scale, the total and active parameter counts across our selected architectures are not perfectly aligned. Although we explicitly account for this limitation in our design rationale, we cannot fully disentangle the effects of architectural family and model capacity on observed performance.

\textcolor{black}{Finally, we did not independently revalidate the semantic normalization stage for each model. Hence, the observed variations may partially originate from model-specific interactions with the second formatting prompt rather than from the initial vulnerability assessment only. Future replications should report normalization agreement separately for each model.}

\paragraph*{External Validity}
The most significant threat to external validity is the limited scope of the benchmark dataset. Our study, like the reference work, relies exclusively on the SmartBugs curated dataset~\cite{ferreira2020smartbugs}, which comprises 142 annotated contracts spanning nine DASP Top 10 vulnerability categories. While this dataset is widely used in the literature and enables direct comparison with Chen et al.~\cite{chen2025chatgpt}, its limited size constrains the generalizability of our findings. Results may not generalize to synthetic datasets~\cite{iuliano2025automated} or to real-world audit scenarios involving more diverse, complex, or more recent contract codebases.

\textcolor{black}{A further threat concerns possible benchmark contamination. SmartBugs-curated is a public and widely used benchmark that predates the evaluated models. Some contracts, annotations, or derivative discussions may have been included in pretraining or instruction-tuning data. Removing comments from the source code prevents direct label leakage during inference, but it does not preclude prior exposure to the benchmark, which may differ across model providers and families, potentially contributing to the observed differences. Since the training corpora of most evaluated models are not fully disclosed, we cannot quantify or rule out this threat.}

\paragraph*{Construct Validity}
A related concern involves the quality and consistency of ground-truth annotations in the SmartBugs curated dataset. The dataset relies on developer-provided comments to identify vulnerability locations and types. If these annotations contain errors, omissions, or ambiguities, the resulting metrics may misrepresent the model's actual performance. We adopted the dataset as-is to preserve comparability with Chen et al.~\cite{chen2025chatgpt}, but this design choice necessarily inherits any annotation noise present in the original benchmark.

Moreover, excluding reasoning models from our architectural comparison, while methodologically justified to avoid confounding the Chain-of-Thought prompt pattern with the model's internal reasoning mechanism, limits the construct coverage of our study. Reasoning-oriented architectures are an increasingly prominent paradigm in the LLM landscape, and excluding them means that our findings speak only to a subset of the architectural space relevant to practitioners.

\paragraph*{Conclusion Validity}
A primary threat to conclusion validity is the statistical power of our experimental design. With only five repetitions per experiment and a relatively small dataset of 142 contracts, the confidence intervals around our performance estimates are wide, particularly for vulnerability classes with few positive instances, limiting our ability to draw statistically robust conclusions about inter-architecture differences, especially when observed F1-score gaps are small. In particular, we do not report bootstrap confidence intervals or formal significance tests on the macro-level comparisons, so the strength of our cross-architecture claims should be interpreted accordingly.

Finally, the selection of representative models per architectural family. Our conclusions about each architectural family are based on a single model instance. Since models within the same family may differ substantially in behavior due to training data, instruction tuning, and alignment procedures, as evidenced by the divergence between Qwen-3.5 and DeepSeek-V2-Lite, a single representative may not adequately characterize the family as a whole. Our findings should therefore be interpreted as evidence about the specific models evaluated rather than as definitive characterizations of their respective architectural families.

\section{Conclusion and Future Work}
\label{sec:conclusion}
This paper revisits the empirical study by Chen et al.~\cite{chen2025chatgpt} on LLM-based smart contract vulnerability detection to verify whether their findings hold under changes in the underlying architecture.
Our results show that the detection profile reported in the original study does not transfer uniformly across architectures. Outside the dense Transformer family, the same prompt elicits qualitatively different regimes, from aggressive over-flagging to markedly conservative behavior. The same qualitative dense-Transformer pattern appears in both GPT-4 and Gemma-3 despite a large-scale gap, which weakens the interpretation that the original result is driven only by model size. Finally, behavior within a single family is not fully predictable from architectural membership alone, as the contrast between the two MoE representatives illustrates.

Several directions for future work emerge. First, our analysis considers a single representative model per architectural family. Expanding this to small cohorts would allow direct measurement of intra-family variability rather than inferring it from isolated cases. Second, the choice of benchmark deserves further attention. Although widely used, the SmartBugs dataset is relatively small and somewhat outdated; repeating the same protocol on more recent corpora would help assess whether the observed architectural effects persist under distributional shifts. Third, the limited number of runs and contract calls necessitate statistical reinforcement of the comparisons, for example, by computing bootstrap confidence intervals for macro-level metrics. Fourth, reasoning-oriented models introduce an additional complication: their internal deliberation mechanisms are not fully compatible with the Chain-of-Thought prompting strategy adopted in our experiments and therefore require a dedicated evaluation setup. Finally, the difficulties we encountered with State Space Models in our preliminary exploration suggest that a focused investigation of this family, disentangling architectural mismatch from the current state of instruction-tuned SSM checkpoints, is worthwhile.

\balance

\section*{Acknowledgment}
Gerardo Iuliano is funded by the European Union – Next Generation EU, Mission 4 Component 2, CUP D42B24002220004, by the Italian Government (MUR), and by AstraKode S.r.l.

\bibliographystyle{IEEEtran}
\bibliography{main.bib}

@article{chen2025chatgpt,
  title={When chatgpt meets smart contract vulnerability detection: How far are we?},
  author={Chen, Chong and Su, Jianzhong and Chen, Jiachi and Wang, Yanlin and Bi, Tingting and Yu, Jianxing and Wang, Yanli and Lin, Xingwei and Chen, Ting and Zheng, Zibin},
  journal={ACM Transactions on Software Engineering and Methodology},
  volume={34},
  number={4},
  pages={1--30},
  year={2025},
  publisher={ACM New York, NY}
}

@article{lo2024large,
  title={Large language models for software engineering: A systematic literature review},
  author={Hou, Xinyi and Zhao, Yanjie and Liu, Yue and Yang, Zhou and Wang, Kailong and Li, Li and Luo, Xiapu and Lo, David and Grundy, John and Wang, Haoyu},
  journal={ACM Transactions on Software Engineering and Methodology},
  volume={33},
  number={8},
  pages={1--79},
  year={2024},
  publisher={ACM New York, NY}
}

@inproceedings{ferreira2020smartbugs,
  title={Smartbugs: A framework to analyze solidity smart contracts},
  author={Ferreira, Jo{\~a}o F and Cruz, Pedro and Durieux, Thomas and Abreu, Rui},
  booktitle={Proceedings of the 35th IEEE/ACM international conference on automated software engineering},
  pages={1349--1352},
  year={2020}
}

@misc{wong2018dasp,
  author       = {David Wong and Mason Hemmel},
  title        = {Decentralized Application Security Project: Top 10 of 2018},
  year         = {2018},
  howpublished = {\url{https://dasp.co/index.html}},
}

@inproceedings{de2025llm,
  title={LLM-based generation of solidity smart contracts from system requirements in natural language: the AstraKode case},
  author={De Vito, Gabriele and D'Amici, Damiano and Izzo, Fabiano and Ferrucci, Filomena and Di Nucci, Dario},
  booktitle={2025 IEEE International Conference on Software Analysis, Evolution and Reengineering (SANER)},
  pages={170--180},
  year={2025},
  organization={IEEE}
}

@ARTICLE{9667515,
  author={Kushwaha, Satpal Singh and Joshi, Sandeep and Singh, Dilbag and Kaur, Manjit and Lee, Heung-No},
  journal={IEEE Access}, 
  title={Systematic Review of Security Vulnerabilities in Ethereum Blockchain Smart Contract}, 
  year={2022},
  volume={10},
  number={},
  pages={6605-6621},
  doi={10.1109/ACCESS.2021.3140091}}

@ARTICLE{8847638,
  author={Zou, Weiqin and Lo, David and Kochhar, Pavneet Singh and Le, Xuan-Bach Dinh and Xia, Xin and Feng, Yang and Chen, Zhenyu and Xu, Baowen},
  journal={IEEE Transactions on Software Engineering}, 
  title={Smart Contract Development: Challenges and Opportunities}, 
  year={2021},
  volume={47},
  number={10},
  pages={2084-2106},
  doi={10.1109/TSE.2019.2942301}}

@article{kushwaha,
  title={Systematic review of security vulnerabilities in ethereum blockchain smart contract},
  author={Kushwaha, Satpal Singh and Joshi, Sandeep and Singh, Dilbag and Kaur, Manjit and Lee, Heung-No},
  journal={IEEE Access},
  volume={10},
  pages={6605--6621},
  year={2022},
  publisher={IEEE}
}

@inproceedings{reguard,
  title={Reguard: finding reentrancy bugs in smart contracts},
  author={Liu, Chao and Liu, Han and Cao, Zhao and Chen, Zhong and Chen, Bangdao and Roscoe, Bill},
  booktitle={Proceedings of the 40th International Conference on Software Engineering: Companion Proceeedings},
  pages={65--68},
  year={2018}
}

@inproceedings{mense,
  title={Security vulnerabilities in ethereum smart contracts},
  author={Mense, Alexander and Flatscher, Markus},
  booktitle={Proceedings of the 20th international conference on information integration and web-based applications \& services},
  pages={375--380},
  year={2018}
}

@inproceedings{contractfuzzer,
  title={Contractfuzzer: Fuzzing smart contracts for vulnerability detection},
  author={Jiang, Bo and Liu, Ye and Chan, Wing Kwong},
  booktitle={Proceedings of the 33rd ACM/IEEE international conference on automated software engineering},
  pages={259--269},
  year={2018}
}

@article{sayeed,
  title={Smart contract: Attacks and protections},
  author={Sayeed, Sarwar and Marco-Gisbert, Hector and Caira, Tom},
  journal={Ieee Access},
  volume={8},
  pages={24416--24427},
  year={2020},
  publisher={IEEE}
}

@article{etherfuzz,
  title={EtherFuzz: mutation fuzzing smart contracts for TOD vulnerability detection},
  author={Wang, Xiaoyin and Sun, Jiaze and Hu, Chunyang and Yu, Panpan and Zhang, Bin and Hou, Donghai},
  journal={Wireless Communications and Mobile Computing},
  volume={2022},
  number={1},
  pages={1565007},
  year={2022},
  publisher={Wiley Online Library}
}

@inproceedings{smartscan,
  title={Smartscan: an approach to detect denial of service vulnerability in ethereum smart contracts},
  author={Samreen, Noama Fatima and Alalfi, Manar H},
  booktitle={2021 IEEE/ACM 4th International Workshop on Emerging Trends in Software Engineering for Blockchain (WETSEB)},
  pages={17--26},
  year={2021},
  organization={IEEE}
}

@INPROCEEDINGS{8726833,
  author={Dika, Ardit and Nowostawski, Mariusz},
  booktitle={2018 IEEE International Conference on Internet of Things (iThings) and IEEE Green Computing and Communications (GreenCom) and IEEE Cyber, Physical and Social Computing (CPSCom) and IEEE Smart Data (SmartData)}, 
  title={Security Vulnerabilities in Ethereum Smart Contracts}, 
  year={2018},
  volume={},
  number={},
  pages={955-962},
  doi={10.1109/Cybermatics_2018.2018.00182}}

@article{4,
  title={Do large language models understand us?},
  author={y Arcas, Blaise Ag{\"u}era},
  journal={Daedalus},
  volume={151},
  number={2},
  pages={183--197},
  year={2022},
  publisher={MIT Press One Rogers Street, Cambridge, MA 02142-1209, USA journals-info~…}
}

@article{5,
  title={Language models are unsupervised multitask learners},
  author={Radford, Alec and Wu, Jeffrey and Child, Rewon and Luan, David and Amodei, Dario and Sutskever, Ilya and others},
  journal={OpenAI blog},
  volume={1},
  number={8},
  pages={9},
  year={2019}
}

@article{6,
  title={Language models are few-shot learners},
  author={Brown, Tom and Mann, Benjamin and Ryder, Nick and Subbiah, Melanie and Kaplan, Jared D and Dhariwal, Prafulla and Neelakantan, Arvind and Shyam, Pranav and Sastry, Girish and Askell, Amanda and others},
  journal={Advances in neural information processing systems},
  volume={33},
  pages={1877--1901},
  year={2020}
}

@inproceedings{7,
  title={Bert: Pre-training of deep bidirectional transformers for language understanding},
  author={Devlin, Jacob and Chang, Ming-Wei and Lee, Kenton and Toutanova, Kristina},
  booktitle={Proceedings of the 2019 conference of the North American chapter of the association for computational linguistics: human language technologies, volume 1 (long and short papers)},
  pages={4171--4186},
  year={2019}
}

@article{nie2026large,
  title={Large language diffusion models},
  author={Nie, Shen and Zhu, Fengqi and You, Zebin and Zhang, Xiaolu and Ou, Jingyang and Hu, Jun and Zhou, Jun and Lin, Yankai and Wen, Ji-Rong and Li, Chongxuan},
  journal={Advances in Neural Information Processing Systems},
  volume={38},
  pages={50608--50646},
  year={2026}
}

@article{achiam2023gpt,
  title={Gpt-4 technical report},
  author={Achiam, Josh and Adler, Steven and Agarwal, Sandhini and Ahmad, Lama and Akkaya, Ilge and Aleman, Florencia Leoni and Almeida, Diogo and Altenschmidt, Janko and Altman, Sam and Anadkat, Shyamal and others},
  journal={arXiv preprint arXiv:2303.08774},
  year={2023}
}

@article{touvron2023llama,
  title={Llama: Open and efficient foundation language models},
  author={Touvron, Hugo and Lavril, Thibaut and Izacard, Gautier and Martinet, Xavier and Lachaux, Marie-Anne and Lacroix, Timoth{\'e}e and Rozi{\`e}re, Baptiste and Goyal, Naman and Hambro, Eric and Azhar, Faisal and others},
  journal={arXiv preprint arXiv:2302.13971},
  year={2023}
}

@misc{anthropic2023claude,
  author       = {Anthropic},
  title        = {Introducing Claude},
  year         = {2023},
  howpublished = {\url{https://www.anthropic.com/}},
}

@misc{2025gemini,
      title={Gemini: A Family of Highly Capable Multimodal Models}, 
      author={Gemini Team},
      year={2025},
      eprint={2312.11805},
      archivePrefix={arXiv},
      primaryClass={cs.CL},
      url={https://arxiv.org/abs/2312.11805}, 
}

@article{jiang2024mixtral,
  title={Mixtral of Experts},
  author={Jiang, Albert Q. and Sablayrolles, Alexandre and Roux, Antoine and Mensch, Arthur and Savary, Blanche and Bamford, Chris and Chaplot, Devendra Singh and de las Casas, Diego and Hanna, Emma Bou and Bressand, Florian and Lengyel, Gianna and Bour, Guillaume and Lample, Guillaume and Lavaud, Lélio Renard and Saulnier, Lucile and Lachaux, Marie-Anne and Stock, Pierre and Subramanian, Sandeep and Yang, Sophia and Antoniak, Szymon and Le Scao, Teven and Gervet, Théophile and Lavril, Thibaut and Wang, Thomas and Lacroix, Timothée and El Sayed, William},
  journal={arXiv preprint arXiv:2401.04088},
  year={2024}
}

@article{tay2022efficient,
  title={Efficient transformers: A survey},
  author={Tay, Yi and Dehghani, Mostafa and Bahri, Dara and Metzler, Donald},
  journal={ACM Computing Surveys},
  volume={55},
  number={6},
  pages={1--28},
  year={2022},
  publisher={ACM New York, NY}
}

@article{jaech2024openai,
  title={Openai o1 system card},
  author={Jaech, Aaron and Kalai, Adam and Lerer, Adam and Richardson, Adam and El-Kishky, Ahmed and Low, Aiden and Helyar, Alec and Madry, Aleksander and Beutel, Alex and Carney, Alex and others},
  journal={arXiv preprint arXiv:2412.16720},
  year={2024}
}

@article{guo2025deepseek,
  title={Deepseek-r1: Incentivizing reasoning capability in llms via reinforcement learning},
  author={Guo, Daya and Yang, Dejian and Zhang, Haowei and Song, Junxiao and Wang, Peiyi and Zhu, Qihao and Xu, Runxin and Zhang, Ruoyu and Ma, Shirong and Bi, Xiao and others},
  journal={arXiv preprint arXiv:2501.12948},
  year={2025}
}

@inproceedings{peng2023rwkv,
  title={Rwkv: Reinventing rnns for the transformer era},
  author={Peng, Bo and Alcaide, Eric and Anthony, Quentin and Albalak, Alon and Arcadinho, Samuel and Biderman, Stella and Cao, Huanqi and Cheng, Xin and Chung, Michael and Derczynski, Leon and others},
  booktitle={Findings of the association for computational linguistics: EMNLP 2023},
  pages={14048--14077},
  year={2023}
}

@inproceedings{solgpt,
  title={Solgpt: A gpt-based static vulnerability detection model for enhancing smart contract security},
  author={Zeng, Shengqiang and Zhang, Hongwei and Wang, Jinsong and Shi, Kai},
  booktitle={International Conference on Algorithms and Architectures for Parallel Processing},
  pages={42--62},
  year={2023},
  organization={Springer}
}

@INPROCEEDINGS{gptlens,
  author={Hu, Sihao and Huang, Tiansheng and İlhan, Fatih and Tekin, Selim Furkan and Liu, Ling},
  booktitle={2023 5th IEEE International Conference on Trust, Privacy and Security in Intelligent Systems and Applications (TPS-ISA)}, 
  title={Large Language Model-Powered Smart Contract Vulnerability Detection: New Perspectives}, 
  year={2023},
  volume={},
  number={},
  pages={297-306},
  doi={10.1109/TPS-ISA58951.2023.00044}}

@INPROCEEDINGS{llmsmartsec,
  author={Mothukuri, Viraaji and Parizi, Reza M. and Massa, James L.},
  booktitle={2024 IEEE International Conference on Blockchain (Blockchain)}, 
  title={LLMSmartSec: Smart Contract Security Auditing with LLM and Annotated Control Flow Graph}, 
  year={2024},
  volume={},
  number={},
  pages={434-441},
  doi={10.1109/Blockchain62396.2024.00064}}

@inproceedings{fellmvp,
  title={Fellmvp: An ensemble llm framework for classifying smart contract vulnerabilities},
  author={Luo, Yu and Xu, Weifeng and Andersson, Karl and Hossain, Mohammad Shahadat and Xu, Dianxiang},
  booktitle={2024 IEEE International Conference on Blockchain (Blockchain)},
  pages={89--96},
  year={2024},
  organization={IEEE}
}

@inproceedings{advscanner,
  title={Advscanner: Generating adversarial smart contracts to exploit reentrancy vulnerabilities using llm and static analysis},
  author={Wu, Yin and Xie, Xiaofei and Peng, Chenyang and Liu, Dijun and Wu, Hao and Fan, Ming and Liu, Ting and Wang, Haijun},
  booktitle={Proceedings of the 39th IEEE/ACM International Conference on Automated Software Engineering},
  pages={1019--1031},
  year={2024}
}

@inproceedings{gptscan,
  title={Gptscan: Detecting logic vulnerabilities in smart contracts by combining gpt with program analysis},
  author={Sun, Yuqiang and Wu, Daoyuan and Xue, Yue and Liu, Han and Wang, Haijun and Xu, Zhengzi and Xie, Xiaofei and Liu, Yang},
  booktitle={Proceedings of the IEEE/ACM 46th International Conference on Software Engineering},
  pages={1--13},
  year={2024}
}

@inproceedings{vulnhunt,
  title={VulnHunt-GPT: a Smart Contract vulnerabilities detector based on OpenAI chatGPT},
  author={Boi, Biagio and Esposito, Christian and Lee, Sokjoon},
  booktitle={Proceedings of the 39th ACM/SIGAPP Symposium on Applied Computing},
  pages={1517--1524},
  year={2024}
}

@article{buterin_ethereum,
  title={Ethereum white paper: a next generation smart contract \& decentralized application platform},
  author={Buterin, Vitalik and others},
  journal={First version},
  volume={53},
  year={2014}
}

@inproceedings{attacks,
  title={A survey of attacks on ethereum smart contracts (sok)},
  author={Atzei, Nicola and Bartoletti, Massimo and Cimoli, Tiziana},
  booktitle={Principles of Security and Trust: 6th International Conference, POST 2017, Held as Part of the European Joint Conferences on Theory and Practice of Software, ETAPS 2017, Uppsala, Sweden, April 22-29, 2017, Proceedings 6},
  pages={164--186},
  year={2017},
  organization={Springer}
}

@article{dao_attack,
  title={Understanding a revolutionary and flawed grand experiment in blockchain: the DAO attack},
  author={Mehar, Muhammad Izhar and Shier, Charles Louis and Giambattista, Alana and Gong, Elgar and Fletcher, Gabrielle and Sanayhie, Ryan and Kim, Henry M and Laskowski, Marek},
  journal={Journal of Cases on Information Technology (JCIT)},
  volume={21},
  number={1},
  pages={19--32},
  year={2019},
  publisher={IGI Global}
}

@misc{iulianoSLR,
      title={Smart Contract Vulnerabilities, Tools, and Benchmarks: An Updated Systematic Literature Review}, 
      author={Gerardo Iuliano and Dario {Di Nucci}},
      year={2024},
      eprint={2412.01719},
      archivePrefix={arXiv},
      primaryClass={cs.SE},
      url={https://arxiv.org/abs/2412.01719}, 
}

@article{ramederSLR,
  title={Review of automated vulnerability analysis of smart contracts on Ethereum},
  author={Rameder, Heidelinde and Di Angelo, Monika and Salzer, Gernot},
  journal={Frontiers in Blockchain},
  volume={5},
  pages={814977},
  year={2022},
  publisher={Frontiers Media SA}
}

@article{liu2023pre,
  title={Pre-train, prompt, and predict: A systematic survey of prompting methods in natural language processing},
  author={Liu, Pengfei and Yuan, Weizhe and Fu, Jinlan and Jiang, Zhengbao and Hayashi, Hiroaki and Neubig, Graham},
  journal={ACM computing surveys},
  volume={55},
  number={9},
  pages={1--35},
  year={2023},
  publisher={ACM New York, NY}
}

@misc{openai2023deprecation,
  author       = {{OpenAI}},
  title        = {{Deprecations - OpenAI API}},
  howpublished = {\url{https://developers.openai.com/api/docs/deprecations}},
  note         = {Accessed: 2026}
}

@misc{openai2023api,
  author       = {{OpenAI}},
  title        = {{GPT - OpenAI API}},
  howpublished = {\url{https://developers.openai.com/api/docs/guides/text}},
  note         = {Accessed: 2026}
}

@misc{openai2023gpt35,
  author       = {{OpenAI}},
  title        = {{GPT-3.5 - OpenAI API}},
  howpublished = {\url{https://developers.openai.com/api/docs/models/gpt-3.5-turbo}},
  note         = {Accessed: 2026}
}

@article{openai2024gpt4o,
  title={Gpt-4o system card},
  author={Hurst, Aaron and Lerer, Adam and Goucher, Adam P and Perelman, Adam and Ramesh, Aditya and Clark, Aidan and Ostrow, AJ and Welihinda, Akila and Hayes, Alan and Radford, Alec and others},
  journal={arXiv preprint arXiv:2410.21276},
  year={2024}
}

@misc{openai2023gpt4,
  author       = {{OpenAI}},
  title        = {{GPT-4 Technical Report}},
  year         = {2023},
  howpublished = {\url{https://arxiv.org/abs/2303.08774}},
  note         = {arXiv:2303.08774},
}

@book{antonopoulos2018masteringSc,
  title={Mastering ethereum: building smart contracts and dapps},
  author={Antonopoulos, Andreas M and Wood, Gavin},
  year={2018},
  publisher={O'reilly Media}
}

@article{szabo1994smartcontract,
  title={Smart contracts http://www. fon. hum. uva. nl/rob/Courses/InformationInSpeech/CDROM/Literature},
  author={Szabo, Nick},
  journal={LOTwinterschool2006/szabo. best. vwh. net/smart. contracts. html},
  year={1994}
}

@book{dannen2017introSolidity,
  title={Introducing Ethereum and solidity},
  author={Dannen, Chris},
  volume={1},
  year={2017},
  publisher={Springer}
}

@inproceedings{kaushal2021immutable,
  title={Immutable smart contracts on blockchain technology: Its benefits and barriers},
  author={Kaushal, Rajesh Kumar and Kumar, Naveen and Panda, Surya Narayan and Kukreja, Vinay},
  booktitle={2021 9th International Conference on Reliability, Infocom Technologies and Optimization (Trends and Future Directions)(ICRITO)},
  pages={1--5},
  year={2021},
  organization={IEEE}
}

@article{politou2019blockMutability,
  title={Blockchain mutability: Challenges and proposed solutions},
  author={Politou, Eugenia and Casino, Fran and Alepis, Efthimios and Patsakis, Constantinos},
  journal={IEEE Transactions on Emerging Topics in Computing},
  volume={9},
  number={4},
  pages={1972--1986},
  year={2019},
  publisher={IEEE}
}

@INPROCEEDINGS{vyper,
  author={Kaleem, Mudabbir and Mavridou, Anastasia and Laszka, Aron},
  booktitle={2020 2nd Conference on Blockchain Research \& Applications for Innovative Networks and Services (BRAINS)}, 
  title={Vyper: A Security Comparison with Solidity Based on Common Vulnerabilities}, 
  year={2020},
  volume={},
  number={},
  pages={107-111},
  doi={10.1109/BRAINS49436.2020.9223278}
}

@article{Rameder2022,
    author = {Rameder, Heidelinde and Di Angelo, Monika and Salzer, Gernot},
    year = {2022},
    month = {03},
    pages = {},
    title = {Review of Automated Vulnerability Analysis of Smart Contracts on Ethereum},
    volume = {5},
    journal = {Frontiers in Blockchain},
    doi = {10.3389/fbloc.2022.814977}
}

@article{li2024static,
  title={Static Application Security Testing (SAST) Tools for Smart Contracts: How Far Are We?},
  author={Li, Kaixuan and Xue, Yue and Chen, Sen and Liu, Han and Sun, Kairan and Hu, Ming and Wang, Haijun and Liu, Yang and Chen, Yixiang},
  journal={Proceedings of the ACM on Software Engineering},
  volume={1},
  number={FSE},
  pages={1447--1470},
  year={2024},
  publisher={ACM New York, NY, USA}
}

@article{vidal2024openscv,
  title={Openscv: An open hierarchical taxonomy for smart contract vulnerabilities},
  author={Vidal, Fernando Richter and Ivaki, Naghmeh and Laranjeiro, Nuno},
  journal={Empirical Software Engineering},
  volume={29},
  number={4},
  pages={101},
  year={2024},
  publisher={Springer}
}

@article{iuliano2026smart,
title = {Smart contract vulnerabilities, tools, and benchmarks: An updated systematic literature review},
journal = {Journal of Systems and Software},
volume = {236},
pages = {112788},
year = {2026},
issn = {0164-1212},
doi = {https://doi.org/10.1016/j.jss.2026.112788},
url = {https://www.sciencedirect.com/science/article/pii/S0164121226000221},
author = {Gerardo Iuliano and Dario {Di Nucci}}
}

@inproceedings{iuliano2025automated,
  title={Automated vulnerability injection in solidity smart contracts: A mutation-based approach for benchmark development},
  author={Iuliano, Gerardo and Allocca, Luigi and Cicalese, Matteo and Di Nucci, Dario},
  booktitle={Proceedings of the 29th International Conference on Evaluation and Assessment in Software Engineering},
  pages={91--101},
  year={2025}
}

@article{10.1145/3687251.3687253,
author = {Boi, Biagio and Esposito, Christian and Lee, Sokjoon},
title = {Smart Contract Vulnerability Detection: The Role of Large Language Model (LLM)},
year = {2024},
issue_date = {June 2024},
publisher = {Association for Computing Machinery},
address = {New York, NY, USA},
volume = {24},
number = {2},
issn = {1559-6915},
url = {https://doi.org/10.1145/3687251.3687253},
doi = {10.1145/3687251.3687253},
journal = {SIGAPP Appl. Comput. Rev.},
month = aug,
pages = {19–29},
numpages = {11}
}

@article{ding2025smartguard,
  title={SmartGuard: An LLM-enhanced framework for smart contract vulnerability detection},
  author={Ding, Hao and Liu, Yizhou and Piao, Xuefeng and Song, Huihui and Ji, Zhenzhou},
  journal={Expert Systems with Applications},
  volume={269},
  pages={126479},
  year={2025},
  publisher={Elsevier}
}

@article{wei2024llm,
  title={Llm-smartaudit: Advanced smart contract vulnerability detection},
  author={Wei, Zhiyuan and Sun, Jing and Zhang, Zijiang and Zhang, Xianhao and Li, Meng and Hou, Zhe},
  journal={arXiv preprint arXiv:2410.09381},
  year={2024}
}

@inproceedings{dao2024transformers,
  title={Transformers are SSMs: generalized models and efficient algorithms through structured state space duality},
  author={Dao, Tri and Gu, Albert},
  booktitle={Proceedings of the 41st International Conference on Machine Learning},
  pages={10041--10071},
  year={2024}
}

@inproceedings{ghattas2025pruning,
  title={On pruning state-space llms},
  author={Ghattas, Tamer and Hassid, Michael and Schwartz, Roy},
  booktitle={Proceedings of the 2025 Conference on Empirical Methods in Natural Language Processing},
  pages={18811--18825},
  year={2025}
}

@inproceedings{zhanrouting,
  title={Routing Mamba: Scaling State Space Models with Mixture-of-Experts Projection},
  author={Zhan, Zheng and Ren, Liliang and Wang, Shuohang and Liu, Liyuan and Liu, Yang and Gong, Yeyun and Wang, Yanzhi and others},
  booktitle={The Thirty-ninth Annual Conference on Neural Information Processing Systems}
}

@inproceedings{wangunderstanding,
  title={Understanding and Mitigating Bottlenecks of State Space Models through the Lens of Recency and Over-smoothing},
  author={Wang, Peihao and Cai, Ruisi and Wang, Yuehao and Zhu, Jiajun and Srivastava, Pragya and Wang, Zhangyang and Li, Pan},
  booktitle={The Thirteenth International Conference on Learning Representations}
}

@article{patro2025mamba,
  title={Mamba-360: Survey of state space models as transformer alternative for long sequence modelling: Methods, applications, and challenges},
  author={Patro, Badri Narayana and Agneeswaran, Vijay Srinivas},
  journal={Engineering Applications of Artificial Intelligence},
  volume={159},
  pages={111279},
  year={2025},
  publisher={Elsevier}
}

@book{wohlin2012experimentation,
  title={Experimentation in software engineering},
  author={Wohlin, Claes and Runeson, Per and H{\"o}st, Martin and Ohlsson, Magnus C and Regnell, Bj{\"o}rn and Wessl{\'e}n, Anders and others},
  volume={236},
  year={2012},
  publisher={Springer}
}

@article{ruopp2008youden,
  title={Youden Index and optimal cut-point estimated from observations affected by a lower limit of detection},
  author={Ruopp, Marcus D and Perkins, Neil J and Whitcomb, Brian W and Schisterman, Enrique F},
  journal={Biometrical Journal: Journal of Mathematical Methods in Biosciences},
  volume={50},
  number={3},
  pages={419--430},
  year={2008},
  publisher={Wiley Online Library}
}

\end{document}